\documentclass[aps,amsfonts,pre,twocolumn,superscriptaddress,showpacs,eqsecnum]{revtex4-1}

\usepackage[normalem]{ulem}
\usepackage{float}
\usepackage[usenames]{color}
\usepackage{graphicx}
\usepackage{ulem}
\usepackage{amsfonts}
\usepackage{amsmath}
\usepackage{natbib}
\usepackage[caption=false]{subfig} 
\usepackage{epsfig,amsmath,amssymb,bm,epsf,psfrag,verbatim}
\usepackage{units} 

\def\Lr{L_{\textrm{rest}}}
\def\Le{L_0}
\def\sr{g}
\def\br{\kappa}
\def\t{\tau}
\newcommand{\bpi}{\boldsymbol{\pi}}
\newcommand{\deriv}[2]{\frac{d #1}{d #2}}

\begin{document}

\title{Finite-temperature buckling of an extensible rod}

\author{Deshpreet Singh Bedi}
\affiliation{Department of Physics, University of Michigan, Ann Arbor, MI 48109, USA }

\author{Xiaoming Mao}
\affiliation{Department of Physics, University of Michigan, Ann Arbor, MI 48109, USA }

\date{\today}
\begin{abstract}
Thermal fluctuations can play an important role in the buckling of elastic objects at small scales, such as polymers or nanotubes.  In this paper, we study the finite-temperature buckling transition of an extensible rod by analyzing fluctuation corrections to the elasticity of the rod.  We find that, in both two and three dimensions, thermal fluctuations delay the buckling transition, and near the transition, there is a critical regime in which fluctuations are prominent and make a contribution to the effective force that is of order $\sqrt{T}$.  We verify our theoretical prediction of the phase diagram with Monte Carlo simulations.

\end{abstract}


\maketitle

\section{Introduction}
When a thin elastic rod is under compression on its two ends, it experiences an instability towards buckling as the compression exceeds a critical value; this is the classical Euler buckling problem~\cite{Euler1744,Landau1986}.  This critical compression is determined by the competition between the compression and bending energy costs of the rod.  The buckling instability plays an important role in many problems in fields ranging from physics to engineering and biology~\cite{Jones2006,Harris1984,Kucken2004,Mullin2007,Das2008,Broedersz2014}.

More recently, experimental studies on buckling phenomena at small length scales, such as the buckling of stiff or semiflexible polymers, nano-filaments, and nanotubes, have been enabled by advances in various technologies~\cite{Kovar2004,Dogterom1997,Brangwynne2006,Ryu2009,Kuzumaki2006}.  These studies may lead to novel devices that utilize transitions between multiple mechanical ground states.  At these small scales, it is necessary to include effects of thermal fluctuations, which have been shown to lead to interesting phenomena near mechanical instabilities in various systems~\cite{Mao2015,Zhang2015,Mao2013,Mao2013a,Rocklin2014,Dennison2013,Bowick2009}.  Such thermal-fluctuation effects have been theoretically investigated, and phenomena such as corrections to the projected end-to-end length, shifts in the critical compression, and softening of the buckling transition have been discovered~\cite{Odijk1998,Hansen1999,Carr2001,Lawrence2002,Baczynski2007,Blundell2009,Emanuel2007,Golubovic2000}.  However, most of these theoretical studies on how fluctuations renormalize the buckling transition have focused on the case of inextensible polymers and have employed the worm-like chain model, which assumes that the polymer has a \emph{constant contour length}.  This is an idealized limit where the rod cannot be stretched/compressed.  For real rods, although the resistance against stretching is much stronger than that against bending, it is worthwhile to discuss whether the extensibility of the rod changes what is known about buckling at finite temperature.

\begin{figure*}
	\centering
		\subfloat{\includegraphics[width=.48\textwidth]{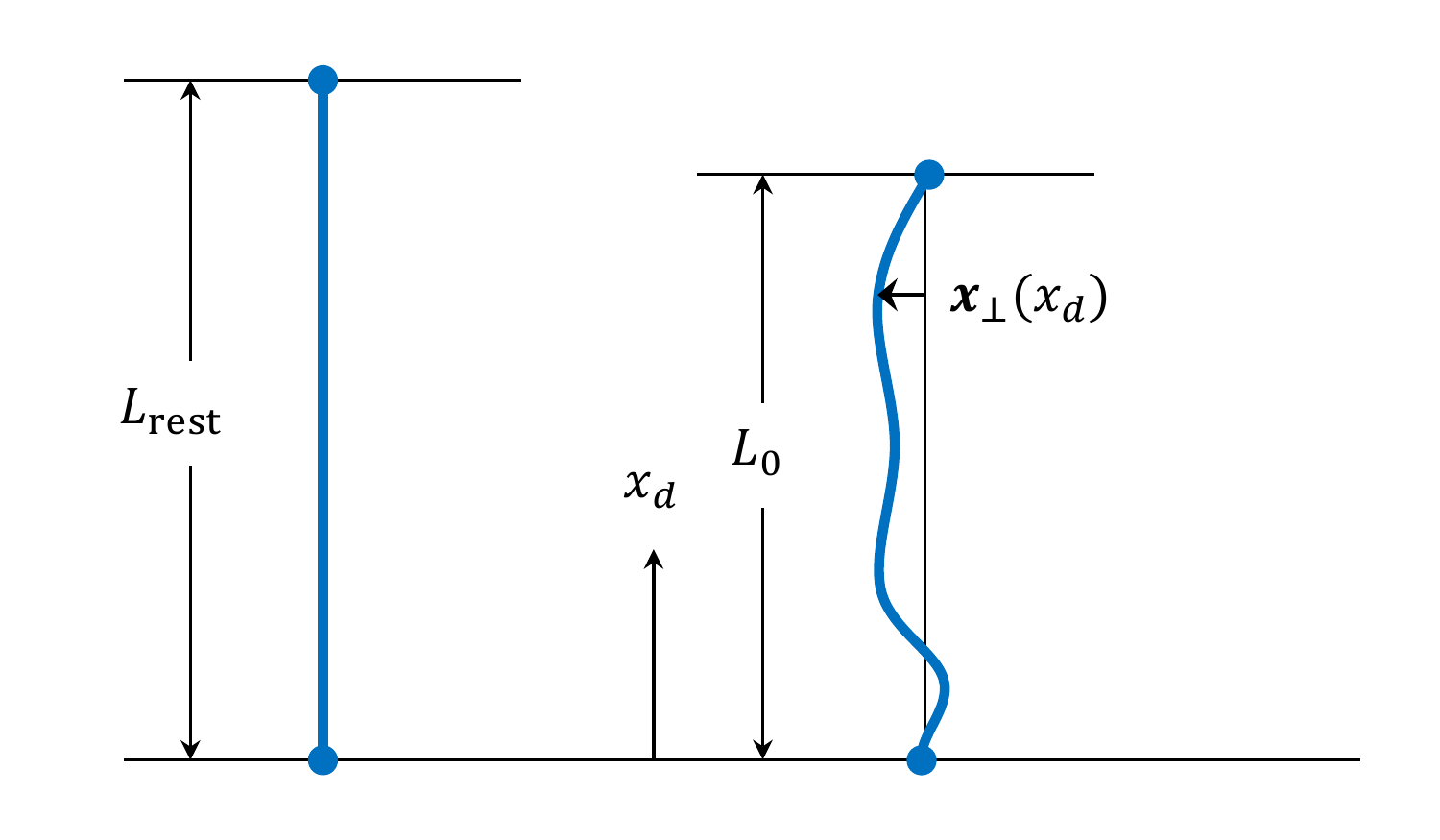}}
		\subfloat{\includegraphics[width=.48\textwidth]{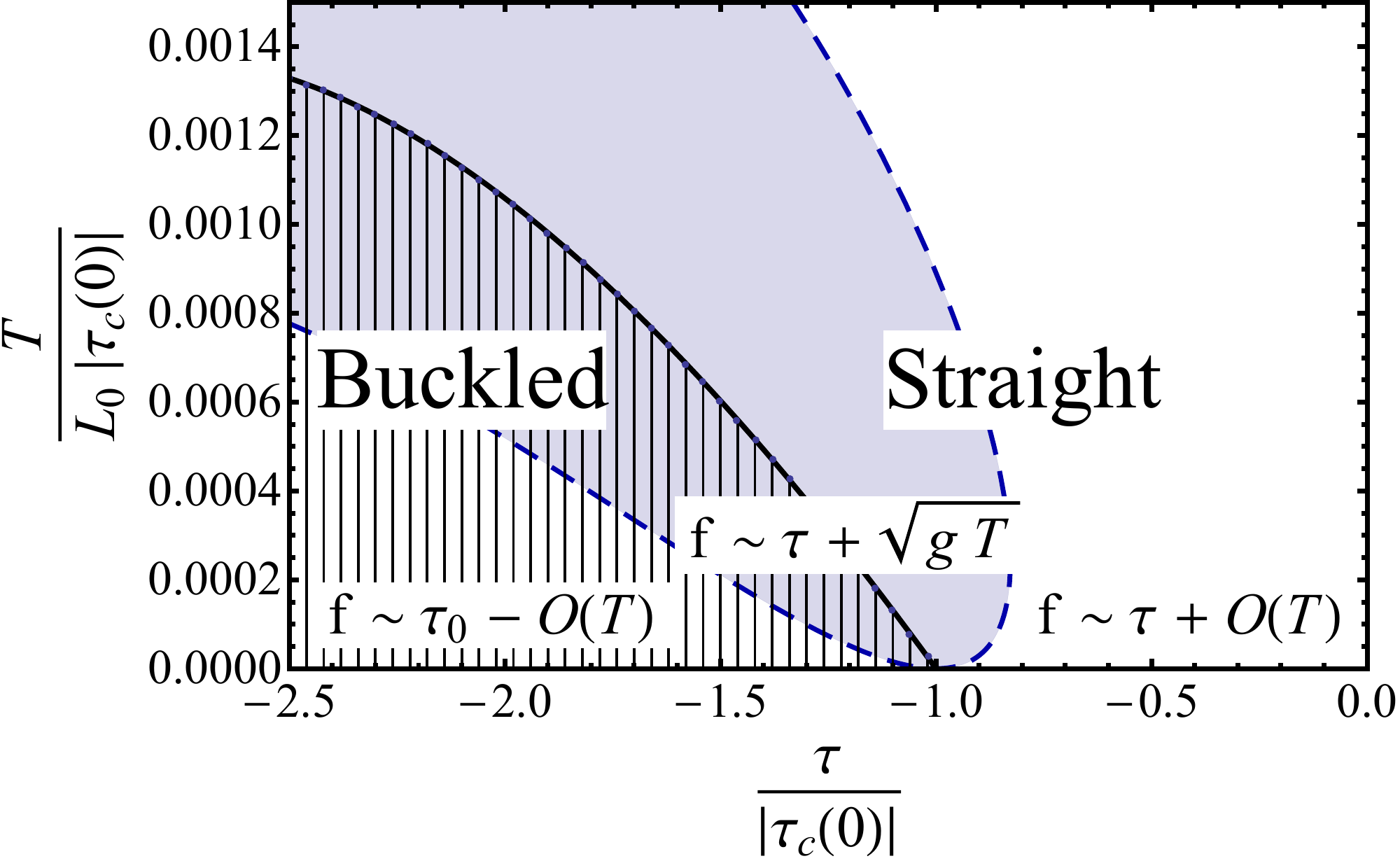}}
	\caption{(Left) Illustration of an extensible rod under compression. (Right) Predicted phase diagram for two-dimensional finite-temperature buckling of an extensible rod in the plane of normalized compression and temperature (as defined in Sec.~\ref{SEC:Model}).  The thick black curve represents the phase boundary between the straight (unshaded) and the buckled (shaded with vertical lines) phases.  The dashed curve denotes the boundary of the critical regime (light blue region) where the thermal-fluctuation correction to the effective force is of $O(\sqrt{T})$.}
	\label{FIG:PhaseDiagFull}
\end{figure*}

In this paper, we investigate finite-temperature buckling using a model elastic energy that allows for rod extensions.  In this model, the end-to-end distance is the control parameter (fixed-strain ensemble), and the rod is allowed to have transverse fluctuations, which both stretch/compress and bend the rod.  By integrating out higher-momentum modes which couple to the first fundamental mode through anharmonic terms, we calculate fluctuation corrections to the rigidity and analyze the buckling transition of the renormalized theory.  We find that, in both two and three dimensions, thermal fluctuations shift the buckling transition to larger-magnitude values of compression.  Our Monte Carlo simulations verify the analytic phase diagram we obtain (Fig.~\ref{FIG:PhaseDiagFull}).  In addition, we also analytically calculate the effective force of the rod, showing that, close to the buckling transition, thermal fluctuations are prominent and contribute an $O(\sqrt{T})$ correction to the effective force.

It is worth pointing out that, in the presence of thermal fluctuations, the rod is never completely ``straight.''  The physical meaning of having a ``straight-buckled'' transition is that the mean-square transverse fluctuations of the rod (e.g., the mean-square transverse displacement of the midpoint of the rod) change from zero in the straight phase to a nonzero value in the buckled phase.  In other words, the elastic free energy minimum of the rod changes from the straight configuration to the buckled configurations.

This paper is organized as follows: we construct the model and discuss the analytic theory in Sec.~\ref{SEC:Model} and present the Monte Carlo simulations in Sec.~\ref{SEC:MC}.  Then, in Sec.~\ref{SEC:Con}, we summarize our results and discuss relations to other studies.

\section{Model and analytic theory}\label{SEC:Model}
\subsection{The extensible-rod Hamiltonian with anharmonic terms}
We consider a thin elastic rod with rest length $\Lr$ embedded in $d$ dimensions, as shown in Fig.~\ref{FIG:PhaseDiagFull}a.  Here, $\Le$ is the end-to-end distance, or projected length, of the rod, which is the control parameter of our theory, and $L$ is the instantaneous contour length in the presence of thermal fluctuations.  

Assuming that the rod is made of a homogeneous material with Young's modulus $E$, its stretching rigidity $\sr$ and bending rigidity $\br$ are given by
\begin{align}\label{EQ:Rod}
	\sr &= \pi a^2 E/\Lr , \nonumber\\
	\br &= EI=(\pi/4) E a^4 ,
\end{align}
where $a$ is the radius of the rod, and $I$ is the moment of inertia of the cross-section. 
Throughout this paper, we require that the relative strengths of the (mechanical) rigidities against bending and against stretching/compression of the rod satisfy
\begin{align}\label{EQ:SBLimit}
	\frac{k_{\perp}}{k_{\parallel}} \propto \frac{\br }{\sr \Lr^3} \propto 
	\left(\frac{a}{\Lr} \right)^2 \ll 1 ,
\end{align}
meaning that it is much more energetically costly to stretch/compress the rod than it is to bend the rod. This is satisfied by most microscopic rod-like objects, including polymers, nanowires and nanotubes~\cite{Emanuel2007}.

The instantaneous stretching/compression elastic energy of the rod can be written as
\begin{align}\label{EQ:Usc}
	U_{sc} &= \frac{1}{2}\sr (L-\Lr)^2 \nonumber\\
	&=  \frac{1}{2}\sr \big\lbrack (L-\Le) + (\Le - \Lr) \big\rbrack^2 .
\end{align}
We define $\t$ to be the force applied to the ends of the straight rod at $T=0$, when there are no thermal fluctuations (i.e., $L=\Le$):
\begin{align}
	\t \equiv \sr (\Le -\Lr) ,
\end{align}
so that $\t>0$ corresponds to stretching of the rod, while $\t<0$ corresponds to compression.  

We proceed to derive the Hamiltonian of the rod for a given compression $\t$ and an instantaneous fluctuation configuration, which is described by
\begin{align}\label{EQ:rx}
	&\mathbf{r}(x_d) = (x_1,\dots,x_{d-1},x_d) 
	\equiv (\mathbf{x}_{\boldsymbol{\perp}},x_d), \nonumber\\
	& 0 \leq x_d \leq \Le,
\end{align}
where $x_d$ parametrizes the rod using the projected end-to-end distance, $\mathbf{r}(x_d)$ is the position of the rod at $x_d$, and $\mathbf{x}_{\boldsymbol{\perp}}$ denotes the transverse displacement of the rod. We define derivatives
\begin{align}
	\bpi(x_d) \equiv \deriv{\mathbf{x}_{\boldsymbol{\perp}}}{x_d} = \left(\deriv{x_1}{x_d},\cdots,\deriv{x_{d-1}}{x_d} \right),
\end{align}
where $\bpi$ is a one- (two-) dimensional vector for the case of a rod embedded in two (three) dimensions.  

The Hamiltonian can then be written as an expansion up to $O(\bpi^4)$,
\begin{align}\label{EQ:Hami}
	H = H_0 +H_2 +H_4,
\end{align}
where
\begin{align}
	H_0 = \frac{\t^2}{2\sr}
\end{align}
is the energy of the straight rod with no fluctuations ($L=\Le$), 
\begin{align}\label{EQ:H2}
	H_2 = \frac{1}{2}\int^{\Le}_{0} dx_d \left( \t |\bpi|^2 
	+ \br | \bpi'|^2 \right) 
\end{align}
contains terms quadratic in $\bpi$, and
\begin{align}\label{EQ:H4}
	H_4 ={}& \frac{1}{2}\int^{\Le}_{0} dx_d 
	\left[ -\frac{\t}{4} |\bpi|^4 
	- \frac{3\br}{2}|\bpi|^2|\bpi'|^2 
	- \br(\bpi \cdot \bpi')^2 \right] \nonumber\\
	&+ \frac{g}{8}\iint\limits^{\Le}_{0} dx_d dx'_d \; |\bpi(x_d)|^2|\bpi(x'_d)|^2
\end{align}
includes terms quartic in $\bpi$. Here $\bpi'$ is shorthand notation for $d\bpi/dx_d$. This Hamiltonian $H$ includes contributions from both stretching/compression as well as bending of the rod, and the details of its derivation are included in App.~\ref{APP:Hami}.  The last term in $H_4$, coming from $\sr (L-\Le)^2/2$, appear to be nonlocal; however, as we shall see, it simply leads to a $\bpi^4$ term in Fourier space with its momentum sum limited to a special channel.  

Note that this formulation with fixed end-to-end distance is the same as the one used in the classical Euler buckling problem in textbooks~\cite{Landau1986}.  A similar formulation has also been used in Refs.~\cite{Carr2001,Lawrence2002}, which focus on quantum aspects of buckling. Additionally, because we are interested in the case where the rod is much more resistant to stretching than it is to bending, the stretching can be taken to be small and highly homogeneous throughout the rod.  This allows for the approximation to be made that the parameters $\br$ and $g$ are uniform along the rod, as in Ref.~\cite{Odijk1995}.

\subsection{Classical ($T=0$) Euler buckling}
The $T=0$ buckling transition is obtained by analyzing the stability of the quadratic coefficient of the Hamiltonian $H$, while the $T=0$ configuration is determined by the location of the minimum of $H$.  
It is convenient to analyze this Hamiltonian in momentum space.  In order to use the convenient exponential form of the Fourier transform, 
we employ the trick of extending the end-to-end distance of the rod to $x_d \in [-\Le, \Le]$ to obtain periodic boundary conditions from the physical fixed boundary conditions that $\mathbf{x}_{\boldsymbol{\perp}}=0$ at $x_d=0,\Le$ (further discussion of this can be found in App.~\ref{APP:Hami}). The quadratic-order Hamiltonian, which is sufficient to ascertain the stability of the system, can then be written as
\begin{align}
	H_2 &= \frac{1}{8\Le} \sum_{q} (\t + \br q^2) \bpi _q^2,
\end{align}
where
\begin{equation}
  q = \frac{n\pi}{\Le}, \hspace{6pt} n \in \mathbb{Z}\setminus\{0\}.
\end{equation} 
Although the sum seemingly counts excess modes by including both positive and negative values of $q$, these modes are not actually independent: because $\bpi(x_d)$ is real and even, $\bpi_q$ satisfies the constraints that
\begin{equation}
\bpi_q = \bpi_{-q} = \bpi^*_q,
\end{equation}
so that the above sum is even in $q$, and the number of independent modes is the same as in the case of expanding $H$ in terms of $\sin (n\pi x_d/\Le)$.
It is straightforward to extract the $T=0$ Euler buckling condition from this equation.  The magnitude of the lowest allowed momentum mode is $q_1 = \pi/\Le$, 
since the $q=0$ mode is excluded by the above fixed-end boundary conditions.  
In order for the Hamiltonian to have a stable equilibrium at $\bpi_q = \mathbf{0}$, its matrix representation must be positive definite -- all its eigenvalues must be positive:
\begin{align}
	\t + \br q^2 > 0 \;\;\forall q.
\end{align}
Applying this condition to the lowest mode, we obtain the critical compression
\begin{align}
	\t_c(0)= -\br\frac{\pi^2}{\Le^2} ,
\end{align}
where the $0$ in parentheses indicates that this is a $T=0$ result. Recall that $\t < 0$ corresponds to compression of the rod, so that for any compression $\t > \t_c(0)$ (i.e., compression with a magnitude less than that of the critical value), the rod remains straight. 

For $\t<\t_c(0)$, on the other hand, the harmonic-level Hamiltonian is no longer stable at $\bpi=0$. The number of modes that have become unstable depends on the value of $\t$; for $(n+1)^2\t_c(0) < \t < n^2\t_c(0)$, the first $n$ modes are unstable, as each of their coefficients in $H_2$ is negative. Thus, for increasingly negative values of $\t$, it is possible to have various metastable states corresponding to higher orders of buckling; for any value of $\t < \t_c(0)$, however, the most energetically favorable buckled configuration is the $n=1$ mode. In this paper, we will only be concerned with analyzing the instability of the first momentum mode when considering the buckling transition; therefore, our discussion in the buckled phase will be restricted to the case of $4\t_c(0) < \t < \t_c(0)$, since the second mode becomes unstable for $\t < 4\t_c(0)$. In this range of compression values, the new stable state -- corresponding to the $n=1$ buckled phase -- is fixed by the anharmonic terms in $H_4$ with only $\bpi_1$ nonzero.  As detailed in App.~\ref{APP:Buckled}, the stable configuration is described by
\begin{align}
	|\hat{\mathbf{x}}_{\boldsymbol{\perp}}(x_d)| = \frac{1}{\pi} \sqrt{\frac{4\Le (\t_c(0)-\t)}{g}} \sin\left(\frac{\pi x_d}{\Le}\right)
\end{align}
in the $T=0$ buckled phase, where we have applied the limit of stretching stiffness being much greater than bending stiffness [Eq.~\eqref{EQ:SBLimit}] to obtain $g\Le \gg |\t_c(0)|,|\t|$ to simplify the expression.

\subsection{Fluctuation corrections to stability and finite-temperature buckling}
The finite-temperature phases are determined by the minima of the free energy of the rod, which includes entropic contributions.  
At finite temperature, thermal fluctuations excite all modes of the rod, and these fluctuations renormalize the stability of the rod against buckling.  In order to analyze this entropic effect on the buckling transition, at which the first mode $q_1$ becomes unstable, we follow a procedure similar to that of momentum shell renormalization.  We first separate the first modes from the higher-momentum fluctuation modes:
\begin{equation}
\bpi = \bpi_q^< + \bpi_q^>,
\end{equation}
where
\begin{align}
\bpi_q^< &= \begin{cases} \bpi_q & \text{if } |q|= q_1 \\ 0 & \text{if } |q| > q_1 \end{cases}
\nonumber\\
\text{and}\hspace{12pt}
\bpi_q^> &= \begin{cases} 0 & \text{if } |q|= q_1 \\ \bpi_q & \text{if } |q| > q_1 \end{cases}.
\end{align}
It follows that $\bpi(x_d) = \bpi^<(x_d) + \bpi^>(x_d)$. These two components are decoupled in the quadratic Hamiltonian $H_2$, but $H_4$ has cross terms.

The partition function can then be written as
\begin{align}
	Z =& \int \mathcal{D}\bpi^< \mathcal{D}\bpi^> \,
	e^{-\frac{1}{T}\left( H_0+ H_2(\bpi^<)+H_2(\bpi^>)+H_4(\bpi^<,\bpi^>)  \right)} \nonumber\\
	=& \int \mathcal{D}\bpi^< \, e^{-F^<(\bpi^<)/T},
\end{align}
where, in the second line, we define the Landau free energy
\begin{align}\label{EQ:Fl}
	F^<(\bpi^<) =&~H_0+ H_2(\bpi^<) \nonumber\\
	&~~~ - T \ln \int \mathcal{D}\bpi^> \,
	e^{-\frac{1}{T}\left( H_2(\bpi^>)+H_4(\bpi^<,\bpi^>)  \right)} .
\end{align}

This Landau free energy, with all other modes $\bpi^>$ integrated out, determines the finite-temperature stability of the first mode, as discussed in detail in App.~\ref{APP:Reno}.  The resulting $F^<(\bpi^<)$ includes terms quadratic order in $\bpi^<$,
\begin{align}\label{EQ:F2}
	F^<_2 (\bpi^<) &= \frac{1}{8\Le} \sum_{q}^{<} (\tilde{\t} + \tilde{\br} q^2) \bpi _q^2,
\end{align}
with the original elastic parameters replaced by renormalized ones. The renormalized parameters are given by
\begin{align}\label{EQ:Tilde}
	\tilde{\t} &\equiv \t + \mathcal{A}(\bar{\t})\bar{T}|\t_c(0)|\left[ (d-1)\bar\sr + (d+1)\bar\t \right] \nonumber\\
	\tilde{\br} &\equiv \br - \mathcal{A}(\bar{\t})\bar{T} (3d-1)\br ,
\end{align}
where we have defined the dimensionless quantities
\begin{align}
	\bar{T}&\equiv \frac{T}{\Le |\t_c(0)|}\nonumber\\
	 \bar\sr&\equiv \frac{\sr\Le}{|\t_c(0)|}\nonumber\\
	  \bar\t &\equiv \frac{\t}{\t_c(0)} > 0,
	  \label{EQ:dimlessquants}
\end{align}
and 
\begin{align}
	\mathcal{A}(\bar{\t})\equiv 
	\sum_{n=2}^{\infty} \frac{1}{n^2 - \bar{\t}} 
	= \frac{3\bar{\t} - 1}{2(\bar{\t}-1)\bar{\t}} - \frac{\pi}{2\sqrt{\bar{\t}}}\cot\pi\sqrt{\bar{\t}}.
\end{align}
In accordance with the range of $\t$ we are considering in this paper, $0<\bar{\t}<4$. As will be justified shortly, close to the $T=0$ buckling transition, we can expand $\mathcal{A}(\bar{\t})$ in powers of $\bar{\t}-1$,
\begin{align}\label{EQ:Aexpand}
	\mathcal{A}(\bar{\t}) = \frac{3}{4} +\left(\frac{\pi^2}{12}-\frac{11}{16}  \right) (\bar{\t}-1) 
	+ O[(\bar{\t}-1)^2].
\end{align}

The magnitude of the effective compression, $\tilde{\t}$, as well as the effective bending rigidity, $\tilde{\br}$, both decrease with increasing temperature. It is easy to understand the decrease in $\tilde\t$: thermal fluctuations tend to increase the instantaneous arc length of the rod from its $T=0$ straight-rod length so that the rod effectively feels less compression. In previous works~\cite{Gutjahr2006,Baczynski2007}, the fluctuation correction to $\tilde\br$ was shown to have a prefactor of $(d-2)$ instead of $(3d-1)$ as we have here. The difference arises from the fact that the rod is assumed to be inextensible and, therefore, is  modeled as a worm-like chain in these previous papers, whereas it is extensible in our model. Consequently, it was necessary to reparametrize the rod in terms of $x_d$ rather than the arc length, $s$, modifying the form of the bending energy.

The buckling transition occurs when the first mode becomes unstable, which is when
\begin{align}
	\tilde{\t} + \tilde{\br} q_1^2 = 0.
\end{align}

\begin{figure}[t]
	\centering
		\includegraphics[width=.45\textwidth]{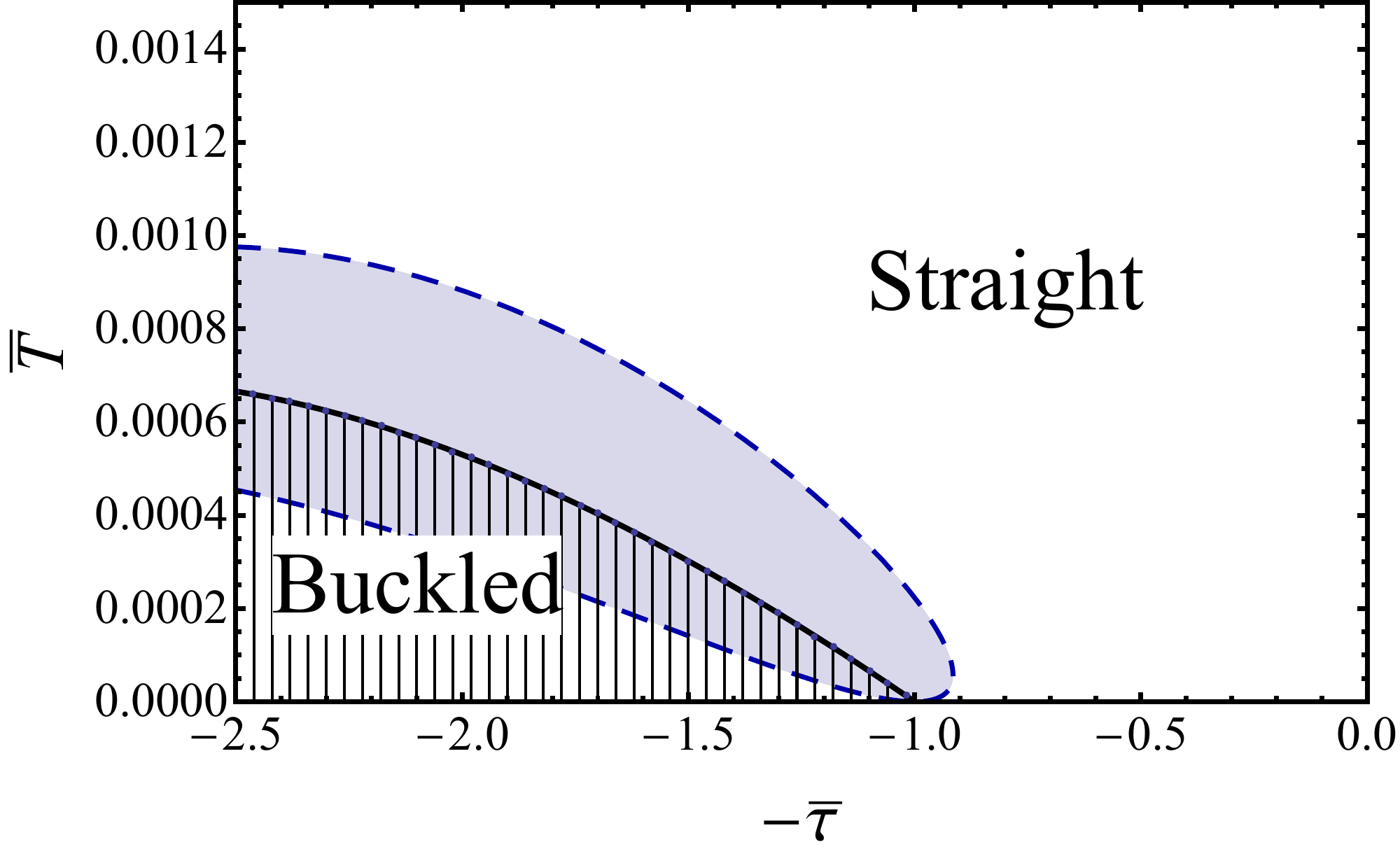}
	\caption{Predicted phase diagram for three-dimensional finite-temperature buckling of an extensible rod in the plane of normalized compression and temperature.  The thick black curve represents the phase boundary between the straight (unshaded) and the buckled (shaded with vertical lines) phases.  The dashed curve denotes the boundary of the critical regime (light blue region) where the thermal-fluctuation correction to the effective force is of $O(\sqrt{T})$.}
	\label{FIG:PhaseDiagrams}
\end{figure}

This condition can be solved to obtain a critical temperature separating the straight ($\bar{T}>\bar{T}_c$) and buckled ($\bar{T}<\bar{T}_c$) phases of the rod for a given compression $\bar\t > 1$,
\begin{equation}
  \bar{T}_c(\bar\t) = \frac{\bar{\t}-1}{\left[(d-1)\bar{\sr} + (d+1)\bar{\t} - (3d-1)\right]      \mathcal{A}(\bar{\t})}.
\end{equation}
The phase boundary in two dimensions determined by this equation is plotted as the solid black line in Fig.~\ref{FIG:PhaseDiagFull}.  The three-dimensional version is shown in Fig.~\ref{FIG:PhaseDiagrams}.  
In the limit that we have been considering of stretching stiffness much greater than bending stiffness [Eq.\eqref{EQ:SBLimit}], we have $\bar{g} \gg 1$, so that we can write a simplified expression for the critical temperature,
\begin{align}\label{EQ:Tc}
	\bar{T}_c (\bar\t) = \frac{\bar\t-1}{(d-1)\bar\sr\mathcal{A(\bar\t)}} \rightarrow \frac{4\left(\bar\t-1\right)}{3(d-1)\bar\sr}.
\end{align}
The expression following the arrow is the limiting case true for sufficiently low temperatures such that $\bar{g}\bar{T}\ll 1$, since, as we can see from the initial equality in Eq.~\eqref{EQ:Tc}, that condition necessitates that $\bar\tau - 1 \ll 1$, as well. In that case, we can write the critical temperature to leading order in $\bar\tau - 1$, allowing us to use the zeroth-order term in the expansion of $\mathcal{A}(\bar{\t})$ in Eq.~\eqref{EQ:Aexpand}.

This leading-order relation can be inverted to obtain an expression for the critical compression for buckling at a finite temperature $T$,
\begin{align}\label{EQ:tauCT}
	\t_c(T) \simeq \t_c(0)\left[1 + \frac{3(d-1) }{4} \bar\sr \bar{T}\right].
\end{align}
This clearly represents a critical compression that is of larger magnitude than the zero-temperature critical value.  In other words, the buckling transition is ``delayed'' by thermal fluctuations.

\subsection{Effective force}
In this paper, we have utilized the ensemble with fixed end-to-end distance $\Le$.  At $T=0$, taking the derivative of the Hamiltonian with respect to $L_0$ yields that the force on the rod is simply $\t = g(\Le - \Lr)$ in the straight phase (with $\t<0$ corresponding to compressional force) and $\t_c(0)$ in the buckled phase (App.~\ref{APP:Buckled}).  

At finite $T$, we determine the effective force $f$ through 
\begin{align}\label{EQ:dFdL}
	f = \frac{\partial F}{\partial \Le} = \frac{\partial \t}{\partial \Le}\frac{\partial F}{\partial \t} = g\frac{\partial F}{\partial \t},
\end{align}
with the free energy given by
\begin{align}\label{EQ:totalF}
	F = -T \ln Z = -T\ln\int \mathcal{D}\bpi^< e^{-F^<(\bpi^<)/T},
\end{align}
where, as defined earlier, $F^<$ is the Landau free energy with only $\bpi^>$ integrated out. It is useful to note that $f$ is calculating by taking the derivative of $F$ with respect to the compression $\t$, rather than by taking the derivative directly with respect to $\Le$.  This is intentional, as the derivative with respect to $\Le$ would also act on the prefactors of $\Le$ in the Fourier transform (or, equivalently, on the integration limits in real space), which would introduce an ultraviolet divergence that scales linearly with the high-momentum cutoff. Strictly speaking, the effective force $f$ obtained via differentiation with respect to $\t$ describes the change of the free energy that occurs with changing the amount of compression $\t$ while keeping $\Le$ constant.


The Landau free energy $F^<$, as defined in Eq.~\eqref{EQ:Fl}, can be written to leading order in $T$ as
\begin{align}\label{EQ:Fl2}
	F^< = H_0 -  T \ln Z_0^> + b_2 T |\bpi_1|^2 +b_4 T |\bpi_1|^4 ,
\end{align}
where 
\begin{align}\label{EQ:Z0}
	Z_0^> = \int \mathcal{D}\bpi^> e^{-H_2(\bpi^>)/T}
\end{align}
is the quadratic-order partition function of $\bpi^>$.  
The coefficients $b_2$ and $b_4$, and the integral over $\bpi_1$, are derived in App.~\ref{APP:EForce}.  We have only needed to retain terms to quadratic order in $\bpi^>$ in \eqref{EQ:Fl2} because $\bpi^>$ modes are stable at $\bpi^>=0$; quartic-order terms in (renormalized) $\bpi_1$ are necessary, however, because the quadratic-order coefficient, $b_2$, can become negative for $\bpi_1$ -- thus, higher-order terms in the potential are needed to evaluate the free energy.  
 
\begin{figure*}[t]
	\centering
		\subfloat{\includegraphics[width=.5\textwidth]{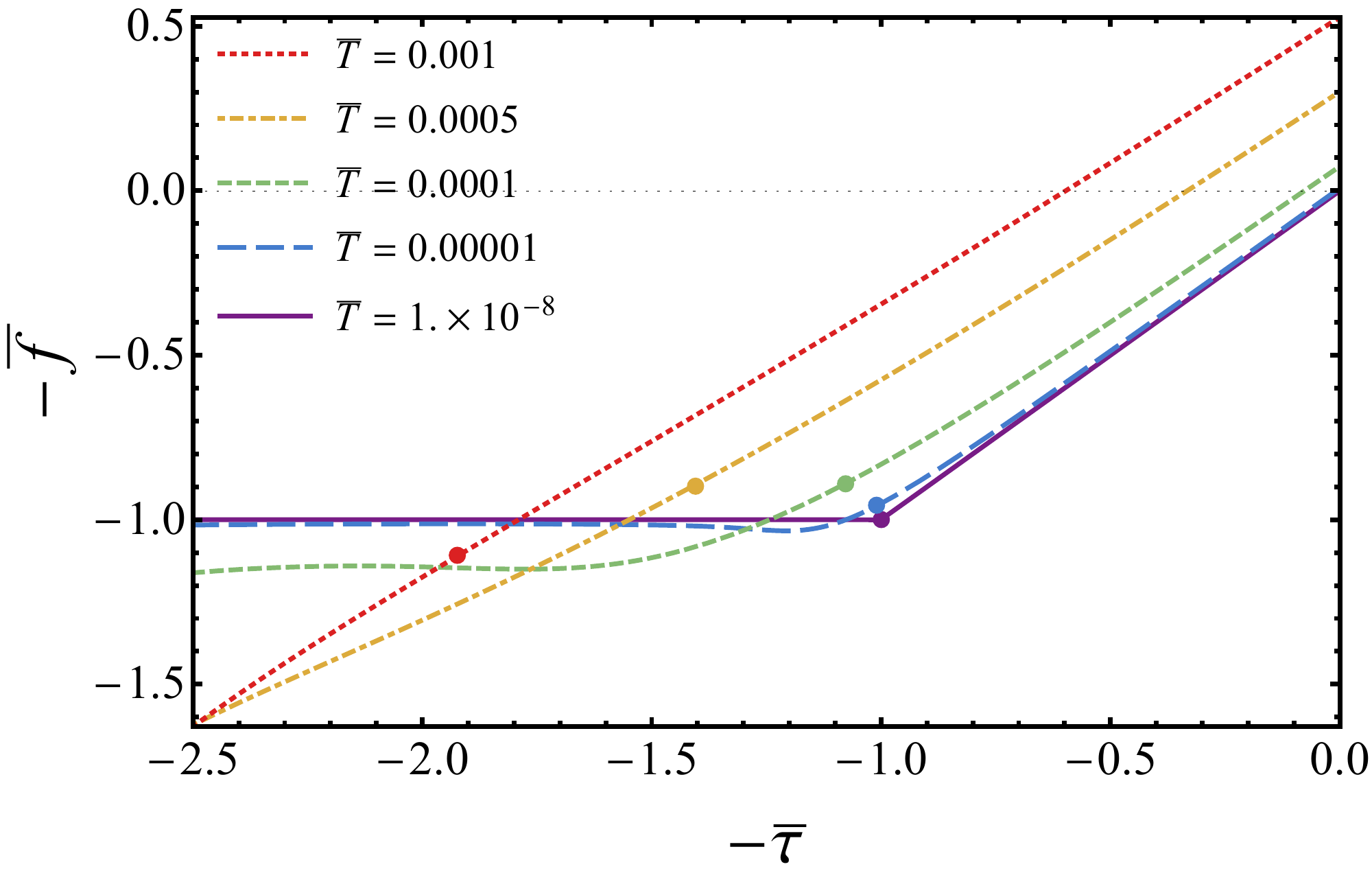}}
		\subfloat{\includegraphics[width=.5\textwidth]{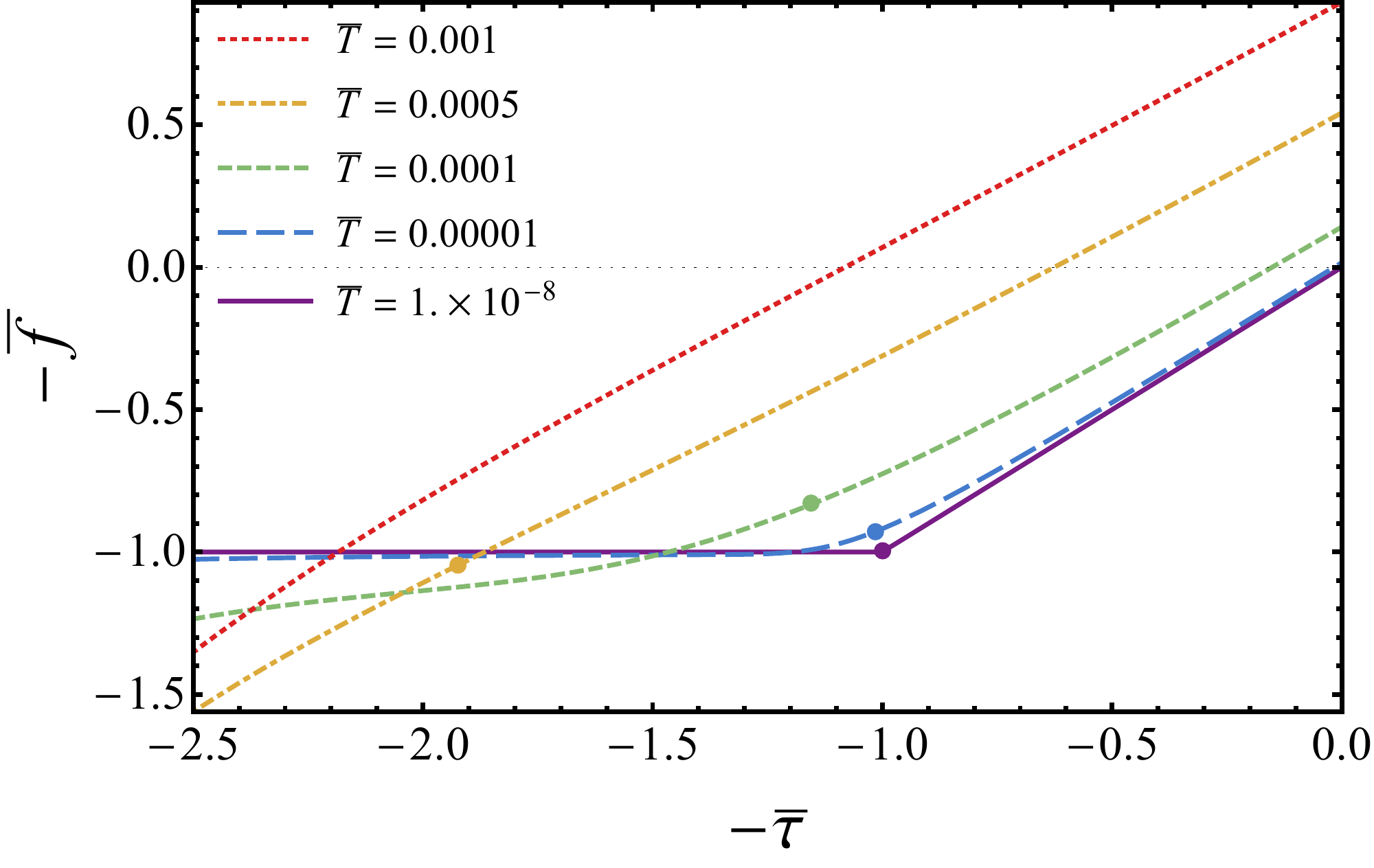}}
	\caption{Plots of the dimensionless effective force in both two (left) and three (right) dimensions at various values of $\bar{T}$. Note that $-\bar{f}=f/|\tau_c(0)|$. Negative values correspond to a compressive force, while positive values correspond to a stretching force. The dots on each curve indicate the transition point between the straight and buckled phases for each value of $\bar{T}$. The $\bar{T} = 0.001$ curve in three dimensions does not have a dot as there is no phase transition at that temperature; the rod remains in the straight phase. 
	}
	\label{FIG:AveForceDiagrams}
\end{figure*}

As detailed in App.~\ref{APP:EForce}, we find that thermal fluctuations reduce the compressional force in the straight phase but enhance it in the buckled phase; these modifications are of order $T$ except very close to the transition for small values of $\bar\sr\bar{T}$, where there is a decrease in the compression of order $\sqrt{T}$, as shown in Fig.~\ref{FIG:AveForceDiagrams}.

\section{Monte Carlo simulations}\label{SEC:MC}
We perform Monte Carlo (MC) simulations in two and three dimensions to corroborate our analytical results.  The rod is discretized into $N$ segments with fixed vertical length $\ell_0 = L_0/N$ along the $x_d$-axis. The segments are allowed to have transverse fluctuations $\mathbf{x}_{\boldsymbol{\perp},j}$ and to, consequently, cause stretching/compression and bending of the rod, as discussed in Sec.~\ref{SEC:Model}.  The fixed boundary conditions necessitate that $\mathbf{x}_{\boldsymbol{\perp},0}=\mathbf{x}_{\boldsymbol{\perp},N}=\mathbf{0}$.  

The Metropolis algorithm is used in our Monte Carlo simulations, in which, at each MC step, a segment is selected at random, and a random trial displacement in the transverse direction is attempted. For a given rod under a certain compression, runs are performed at various temperatures.  We choose the transverse displacement of the middle segment, $ |\mathbf{x}_{\boldsymbol{\perp},\frac{N}{2}}|$, to be our order parameter.  In the straight phase $ |\mathbf{x}_{\boldsymbol{\perp},\frac{N}{2}}| $ is governed by a Gaussian distribution with its mean at $\mathbf{0}$, whereas in the buckled phase, the distribution of $ |\mathbf{x}_{\boldsymbol{\perp},\frac{N}{2}}| $ becomes double-well (in $d=2$) or Mexican-hat ($d=3$) with minima at 
\begin{align}
	 |\hat{\mathbf{x}}_{\boldsymbol{\perp},\frac{N}{2}}|  = \frac{1}{\pi} \sqrt{\frac{4\Le (\t_c(0)-\t)}{g}} .
\end{align}
At the buckling transition, the distribution sharply deviates from Gaussian.  To capture this transition, we calculate the Binder cumulant of the distribution~\cite{Binder1981},
\begin{equation}
U_L = 1 - 
\frac{\langle |\mathbf{x}_{\boldsymbol{\perp},\frac{N}{2}}|^4 \rangle}{3\langle 
|\mathbf{x}_{\boldsymbol{\perp},\frac{N}{2}}|^2 \rangle^2} .
\end{equation}
The value of $U_L$ decreases as the temperature is lowered and the system experiences the straight-to-buckled phase transition.  This decrease becomes increasingly sharp for progressively larger systems, and the simultaneous crossing of Binder cumulant curves for various system sizes determines the location of the critical temperature $T_c$.

To verify our phase diagrams in Figs.~\ref{FIG:PhaseDiagFull} and \ref{FIG:PhaseDiagrams} via the crossing of the Binder cumulant curves, we simulate rods containing 10, 12, and 14 segments (corresponding to $L_0 = 1.0, 1.2, 1.4$, respectively, so that $\ell_0 = 0.1$ is kept fixed).  
As discussed in Sec.~\ref{SEC:Model}, $g=\pi a^2 E/\Lr$, so to keep $a$ and $E$ constant across the various-sized rods (so that each rod has the same cross-section and is made of the same materials), we take the values of $g$ to be $g=10.00, 8.34, 7.15$, corresponding to the three choices of length.  In addition, in accordance with Eqs.~\eqref{EQ:Rod} and~\eqref{EQ:SBLimit}, we take $\br=0.01/\pi^2$.  With these parameters, for $\Le=1$, we have $\t_c(0)=-0.01$. We take $\bar{\t} = \t/\t_c(0)=1.3, 1.5, 1.7$ and vary $T$ to observe the transition.  For these three $\bar{\t}$ values, with $d=2$ and all other elastic parameters corresponding to $\Le = 1$, $T_c=3.78\times 10^{-6}, 6.03 \times 10^{-6}, 8.05 \times 10^{-6}$, satisfying the requirement that the persistence length $l_p \equiv \br/T \simeq 10^{2}$ is much longer than the length of the rod $\Le$, and, therefore, the transverse fluctuations are small.  This justifies the small $\bpi$ expansion we make.


\begin{figure*}[t]
	\centering
	\includegraphics[width=\textwidth]{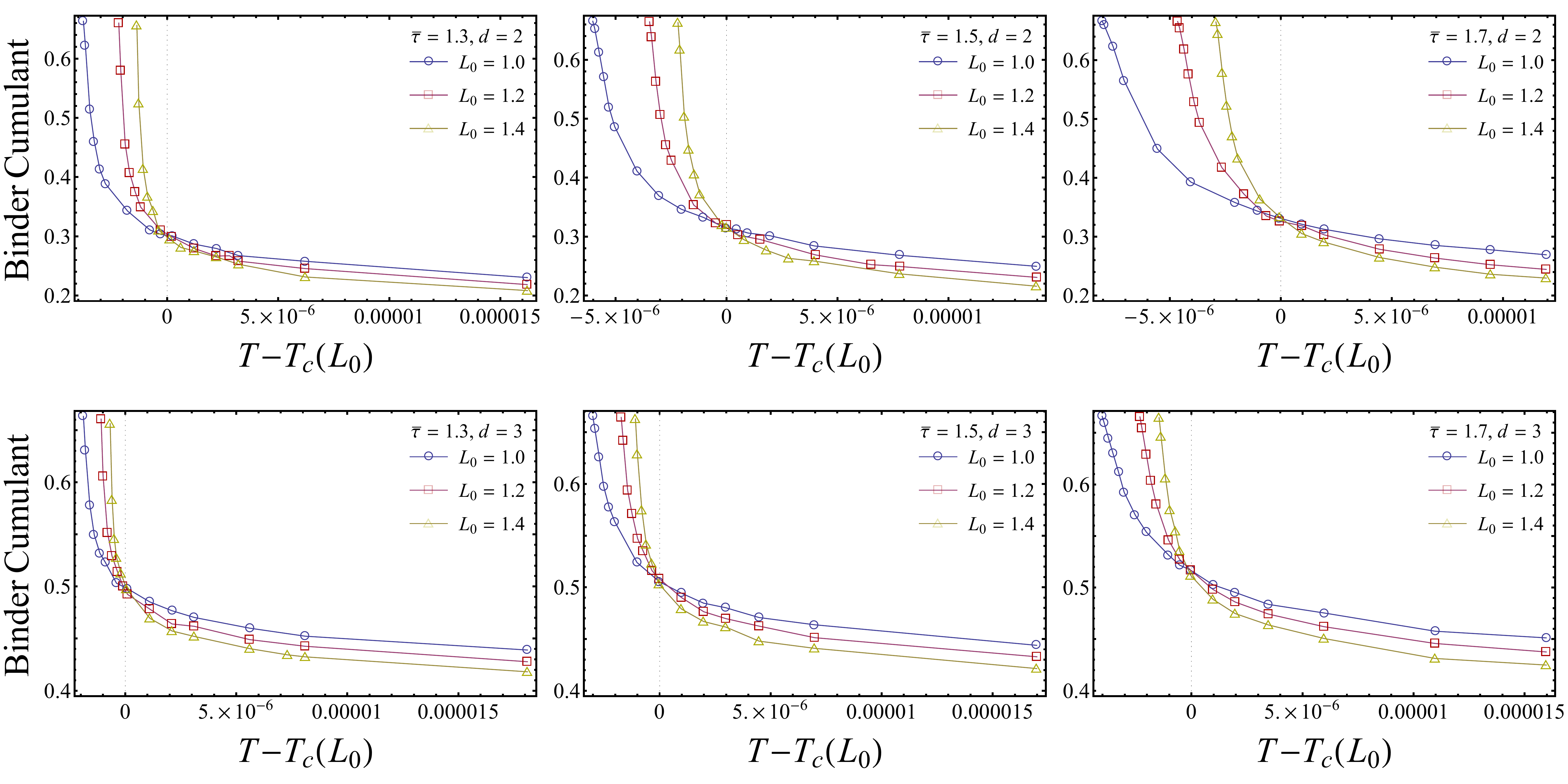}
	\caption{Results of Monte Carlo simulations run with $\bar\t = 1.3,1.5,1.7$ in both two and three dimensions. For each value of $\bar\t$ and in each dimension, three different lengths were simulated. Each data point corresponds to the combined Binder cumulant value of ten independent simulations run with identical parameters. The lines connecting the data points explicitly illustrate that the Binder cumulant curves do indeed simultaneously cross at, or very close to, the respective critical temperatures.}
	\label{FIG:BCPlots}
\end{figure*}

The resulting $U_L$ curves from our MC simulations are shown in Fig.~\ref{FIG:BCPlots}.  Because $T_c$, as given in Eq.~\eqref{EQ:Tc}, depends on the system size $\Le$ through $|\t_c(0)|$, it is necessary to shift the $U_L$ curves by theoretical predictions of $T_c(\t)$ to observe the crossing of the three curves for the different system sizes. The crossing of the three $U_L$ curves for all three values of $\bar{\t}$ in both two and three dimensions verifies our theoretical prediction of the finite-temperature buckling transition.

\section{Conclusion and discussion}\label{SEC:Con}
In this paper, we used both analytic theory and MC simulations to investigate the buckling of an extensible elastic rod at finite temperature.  We find that, in both two and three dimensions, buckling is delayed by thermal fluctuations, and near the transition, there is a critical regime in which the fluctuation correction to the average compression force is of order $\sqrt{T}$.

In comparing the two phase diagrams in Figs.~\ref{FIG:PhaseDiagFull} and \ref{FIG:PhaseDiagrams}, one can observe that the straight-rod phase is more stabilized in three dimensions than in two dimensions. 
This can be intuitively attributed to the fact that in higher dimensions, there are an increasing number of transverse, soft directions in which segments in the straight rod can move compared to the when the rod is buckled. Therefore, the straight rod is increasingly entropically protected, as there are a larger number of accessible states.

Our analytic theory is a perturbative theory that applies to small fluctuations.  This requires that the dimensionless temperature $\bar{T}\ll 1$.  This condition can be written in terms of the persistence length $l_p=\br/T$ as $\Lr/(\pi^2 l_p) \ll 1$, which is satisfied by stiff ($\Lr \ll l_p$) and semiflexible ($\Lr \sim l_p$) polymers.

At lower temperatures, quantum fluctuations also become important.  To make a simple estimate of the temperature scale at which this occurs, we include the kinetic energy term
\begin{align}
	H_{\textrm{kinetic}} = \int dx_d \,\rho \left|\frac{\partial \mathbf{x}_{\boldsymbol{\perp}}}{\partial t}\right|^2 ,
\end{align}
where $\rho$ is the linear mass density of the rod.  
Combining this with the potential energy terms in $H$, we have a phonon energy given by
\begin{align}
	\hbar \omega \sim \frac{\hbar|q|}{\sqrt{\rho}} \sqrt{\t + \br q^2}.
\end{align}
Therefore, in addition to thermal fluctuation corrections, quantum fluctuations also contribute to the renormalization of $\tilde{\t}$ and $\tilde{\br}$, moving the critical $\t$ to a larger compression value (in magnitude) even at $T=0$.  The significance of such contributions from quantum fluctuations can be estimated by comparing $\hbar \omega$ of generic modes with $k_B T$.  For the simple case of stiff polymers of length $10^{-6}\,\text{m}$ and persistence length $10^{-3}\,\text{m}$, we estimate that the characteristic temperature for $\hbar \omega \sim k_B T$ is $T\sim10^{-6}\,\text{K}$, which is extremely low.  Other systems with higher stiffness or shorter lengths may have stronger quantum effects.

Our result that, in both two and three dimensions, the buckling transition is delayed by thermal fluctuations contrasts with previous studies of finite-temperature buckling of polymers using the inextensible worm-like chain model~\cite{Baczynski2007,Emanuel2007}.  
The extensibility of the rod in our model allows for an additional independent quartic-order term in the Hamiltonian, and this term plays an important role in determining the renormalization of the stability of the first mode, leading to the phase diagram shown in Fig.~\ref{FIG:PhaseDiagFull}.

In addition, 
extensive recent studies have focused on zero-temperature mechanical instability in both ordered and disordered systems~\cite{Mao2010,Liu2010,Ellenbroek2011,Mao2011a,Mao2013b,Mao2013c,Zhang2015a,Lubensky2015,Mao2009}, and their behavior at finite temperature remain largely unexplored~\cite{Mao2015,Mao2013,Mao2013a,Rocklin2014,Dennison2013,Bowick1996,Paulose2012}.  Our model provides a clean system which exhibit a shifted second-order transition and the results can be compared to future studies on finite-temperature mechanical instabilities in various systems.

\appendix
\section{Deriving the Hamiltonian of the extensible rod with fluctuations}\label{APP:Hami}
The change in the length of the rod due to thermal fluctuations can be expressed in terms of the $\bpi$ field as
\begin{align}
	L-\Le = \int_0^{\Le} dx_d \; \left(\sqrt{1 + |\bpi|^2} - 1\right) .
\end{align}
Using this, we can then write the stretching/compression elastic energy~\eqref{EQ:Usc} as
\begin{align}
	U_{sc} &= \frac{\t^2}{2\sr}  +
	\t \int_0^{\Le} dx_d \; \left(\sqrt{1 + |\bpi|^2} - 1\right) \nonumber\\
	&\hspace{26pt} + \frac{\sr}{2} \left[ \int_0^{\Le} dx_d \; \left(\sqrt{1 + |\bpi|^2} - 1\right) \right]^2 .
\end{align}

The bending energy of the rod is given by
\begin{align}
U_b = \frac{\br}{2}\int_0^{L} ds \; |d_s \boldsymbol{\hat{\mathbf{t}}}(s)|^2,
\end{align}
where $s$ labels the arc length.  We assume the bending rigidity to be homogeneous along the arc length, given that we are considering the regime where stretching is much more energetically costly than bending.  Here, $\boldsymbol{\hat{\mathbf{t}}}(s)$ is the unit tangent vector at $s$ and $|d_s \boldsymbol{\hat{\mathbf{t}}}(s)|$ is the local curvature.  $U_b$ can also be expressed in terms of $\bpi(x_d)$:
\begin{align}
	U_b = \frac{\br}{2}\int_0^{\Le} dx_d \; \left[\frac{\left| \bpi '\right|^2}{(1 + |\bpi|^2)^{3/2}} - \frac{(\bpi \cdot \bpi')^2}{(1 + |\bpi|^2)^{5/2}} \right] ,
\end{align}
where $\bpi '$ is shorthand for $d \bpi/d x_d$ and we used
\begin{align}
	\boldsymbol{\hat{\mathbf{t}}}(x_d) = \frac{d\mathbf{r}(x_d)/d x_d}{|d\mathbf{r}(x_d)/d x_d|} = \frac{\bpi(x_d)}{|\bpi(x_d)|}
\end{align}
and
\begin{align}
	ds = \sqrt{1 + |\bpi|^2} d x_d .
\end{align}

The total Hamiltonian of the rod is a sum of both the stretching/compression and the bending contributions,
\begin{align}
	H = U_{sc} + U_b .
\end{align}
Expanding this Hamiltonian as a series in $\bpi$ leads to the form in Eq.~\eqref{EQ:Hami}.

To obtain the Fourier transform of this Hamiltonian, we need to pay special attention to the specific boundary conditions of the problem. Here, $\bpi$ has to be a real-valued field, and $\mathbf{x}_{\boldsymbol{\perp}}(x_d)$ (the perpendicular component of $\mathbf{r}$, as defined in Eq.~\ref{EQ:rx}) has to vanish at the two ends, $x_d=0$ and $x_d=\Le$.  This limits the Fourier series of $\mathbf{x}_{\boldsymbol{\perp}}(x_d)$ to $\sin (n\pi x_d/\Le)$ basis functions, and the Fourier series of $\bpi (x_d)$ to $\cos (n\pi x_d/\Le)$ basis functions.  In order to work with the more convenient basis of exponential functions, we necessarily extend the rod to $x_d \in [-\Le, \Le]$ and limit $\bpi(x_d)$ to be real-valued even functions on this interval (correspondingly, $\mathbf{x}_{\boldsymbol{\perp}}(x_d)$ is limited to real-valued odd functions), so that the value of $\bpi(x_d)$ for $-\Le<x_d<0$ is determined by
\begin{align}
	\mathbf{x}_{\boldsymbol{\perp}}(x_d) &= -\mathbf{x}_{\boldsymbol{\perp}}(-x_d) \nonumber\\
	\bpi(x_d) &= \bpi(-x_d).
\end{align}
    Therefore, we can write the Fourier transform as
\begin{align}
	\bpi(x_d) &= \frac{1}{2\Le}\sum_q \bpi_q e^{iqx_d}, \label{eq:piz}\\
	\bpi_q &= \int_{-\Le}^{\Le} dx_d \, \bpi(x_d)e^{-iqx_d},
\end{align}
with 
\begin{equation}
	q = \frac{n\pi}{\Le}, \hspace{6pt} n \in \mathbb{Z}\setminus\{0\}.
\end{equation}
Because $\bpi(x_d)$ is real and even, we have constraints on $\bpi_q$ that
\begin{align}
	\bpi_q = \bpi_{-q} = \bpi_{q}^* .
\end{align}
Therefore, positive and negative $q$ values do not constitute independent modes.

\section{The $T=0$ buckled phase}\label{APP:Buckled}
As we discussed in the main text, for $4\t_c(0)<\t<\t_c(0)$ at $T=0$, the $\bpi=0$ straight state is no longer stable.  The new stable state has $\bpi_1\ne 0$, and the value of $\bpi_1$ is determined by minimizing the total Hamiltonian with both $\bpi^2$ and $\bpi^4$ terms.  Taking $\bpi_q=0$ for all but the first mode ($|q| = \pi/\Le$), the Hamiltonian becomes
\begin{align}
	H = H_0 + \frac{\t-\t_c(0)}{4\Le}|\bpi_1|^2 + 
	\frac{-\frac{3}{2}\t +5\t_c(0) +g\Le}{32 \Le^3} |\bpi_1|^4 .
\end{align}
The minimum-energy configuration is determined by
\begin{align}
	\frac{\partial H}{\partial \bpi_1} \bigg\vert_{\boldsymbol{\hat{\pi}_1}}=0,
\end{align}
where $\hat{\bpi}_1$ denotes the mode corresponding to this minimum-energy configuration. Thus, we find that
\begin{align}\label{EQ:pistar}
	|\hat{\bpi}_1| =\sqrt{\frac{4\Le^2 (\t_c(0)-\t)}{-\frac{3}{2}\t +5\t_c(0) +g\Le}}.
\end{align}
Taking the limit of stretching stiffness much greater than bending stiffness, $g\Le \gg |\t_c(0)|,|\t|$, we obtain
\begin{align}
	|\hat{\bpi}_1| = \sqrt{\frac{4\Le (\t_c(0)-\t)}{g}}.
\end{align}
This leads to the $T=0$ equilibrium buckled configuration
\begin{align}
	|\hat{\mathbf{x}}_{\boldsymbol{\perp}}(x_d)| = \frac{1}{\pi} \sqrt{\frac{4\Le (\t_c(0)-\t)}{g}} \sin\left(\frac{\pi x_d}{\Le}\right).
\end{align}
In two dimensions, where $\mathbf{x}_{\boldsymbol{\perp}}$ is simply a number, there are two degenerate equilibrium buckled configurations corresponding to $\pm |\hat{\mathbf{x}}_{\boldsymbol{\perp}}(x_d)|$. In three dimensions, however, there are an infinite number, consistent with a $U(1)$ symmetry corresponding to rotation about the $x_d$-axis.

The energy of this equilibrium buckled configuration is
\begin{align}
	H =\frac{\t^2}{2g} -\frac{(\t_c(0)-\t)^2}{2g} = \frac{2\t\t_c(0)-\t_c(0)^2}{2g} ,
\end{align}
indicating a constant force at $T=0$ in the buckled phase
\begin{align}
	f = g \frac{\partial H}{\partial \t} = \t_c(0).
\end{align}

\section{Integrating out fluctuations and obtaining the Landau free energy $F^<(\bpi^<)$}\label{APP:Reno}
In this section, we expand the Hamiltonian in terms of $\bpi^<$ and $\bpi^>$ and perform the calculation of integrating out $\bpi^>$.

It is clear that $\bpi^<$ and $\bpi^>$ are decoupled in the quadratic Hamiltonian because they are of different momenta and, therefore, orthogonal, so
\begin{align}
	H_2 = H_2(\bpi^<) + H_2(\bpi^>).
\end{align}
In the quartic-order Hamiltonian $H_4$, on the other hand, they are coupled.  

The partition function of the rod can be written as
\begin{align}
	Z =& \int \mathcal{D}\bpi^< \mathcal{D}\bpi^> \,
	e^{-\frac{1}{T}\left( H_0+ H_2(\bpi^<)+H_2(\bpi^>)+H_4(\bpi^<,\bpi^>)  \right)} \nonumber\\
	={}& Z_0^> \int \mathcal{D}\bpi^< \, e^{-\left( H_0+ H_2(\bpi^<)  \right)/T} \left<e^{-H_4(\bpi^<,\bpi^>) /T}\right>_>,
\end{align}
where $Z_0^>$ is defined in Eq.~\eqref{EQ:Z0} and 
\begin{multline}
	\left<e^{-H_4(\bpi^<,\bpi^>) /T}\right>_>  \\
	 \equiv \frac{1}{Z_0^>} \int \mathcal{D}\bpi^> \, 
	 e^{-\frac{1}{T}\left( H_2(\bpi^>)+H_4(\bpi^<,\bpi^>)  \right)} .
\end{multline}
Following a cumulant expansion, we can then write
\begin{multline}
	\left<e^{-H_4(\bpi^<,\bpi^>) /T}\right>_> \\
	= e^{-\frac{1}{T}\langle H_4 \rangle_> 
	+ \frac{1}{2T^2}\left(\langle H_4^2 \rangle_> - \langle H_4 \rangle_>^2\right) 
	+ \mathcal{O}\left((H_4/T)^3\right) }.
\end{multline}
Since we are ultimately trying to deduce the effect of thermal fluctuations on the stability threshold, we are interested in the corrections to the quadratic terms in $|\bpi^<|$. In the straight phase, $\langle \pi_a \pi_b \rangle \sim T\delta_{ab}$, meaning that $\langle H_4 \rangle \sim T|\bpi^<|^2$ will provide an $O(T)$ correction to the quadratic-order coefficients, while terms from $\langle H_4^2 \rangle/T$ will result in an $O(T^2)$ correction. Since we are doing a perturbative expansion in small fluctuations, which necessitates small temperatures, we need only calculate $\left< H_4\right>_>$.

\begin{widetext}
The explicit form of $H_4$ is given in Eq.~\eqref{EQ:H4}, and here we replace $\bpi$ by $\bpi^< + \bpi^>$. Expanding each term in $H_4$ out, we have
\begin{align}
-\frac{\tau}{8}\int\limits_{0}^{\Le} dx_d \; \langle |\bpi|^4 \rangle_> &= -\frac{\tau}{8}\int\limits_{-L/2}^{L/2} dx_d \; \langle \pi_a\pi_a\pi_b\pi_b \rangle_> \\
\begin{split}
&= -\frac{\tau}{8}\int\limits_{0}^{\Le} dx_d \; \bigg[ \pi_a^<\pi_a^<\pi_b^<\pi_b^< + \pi_a^<\pi_a^<\langle \pi_b^>\pi_b^> \rangle_> + 4\pi_a^<\pi_b^<\langle \pi_a^>\pi_b^> \rangle_> \\
 &\hspace{124pt} + \pi_b^<\pi_b^<\langle \pi_a^>\pi_a^> \rangle_> + \langle \pi_a^>\pi_a^>\pi_b^>\pi_b^> \rangle_> \bigg]
\end{split}\\
&= -\frac{\tau}{8}\int\limits_{0}^{\Le} dx_d \; \left[ |\bpi^<|^4 + 2(d-1)|\bpi^<|^2\frac{1}{(2\Le)^2}\sum_q^> G_{0q} + 4|\bpi^<|^2\frac{1}{(2\Le)^2}\sum_q^> G_{0q} + \langle |\bpi^>|^4 \rangle_> \right] \\
&= -\frac{\tau}{8}\int\limits_{0}^{\Le} dx_d \; \left[ |\bpi^<|^4 + 2(d+1)|\bpi^<|^2\frac{1}{(2\Le)^2}\sum_q^> G_{0q} + \langle |\bpi^>|^4 \rangle_> \right],
\end{align}
\begin{equation}
\begin{split}
-\frac{3\kappa}{4}\int\limits_{0}^{\Le} dx_d \; \langle |\bpi|^2|\partial_{x_d}\bpi|^2 \rangle_> 
= -\frac{3\kappa}{4}\int\limits_{0}^{\Le} dx_d \; \bigg[& |\bpi^<|^2|\partial_{x_d}\bpi^<|^2 + (d-1)|\bpi^<|^2\frac{1}{(2\Le)^2}\sum_q^> q^2 G_{0q} \\& + (d-1)|\partial_{x_d}\bpi^<|^2\frac{1}{(2\Le)^2}\sum_q^> G_{0q} + \langle |\bpi^>|^2|\partial_{x_d}\bpi^>|^2 \rangle_> \bigg],
\end{split}
\end{equation}
\begin{equation}
\begin{split}
-\frac{\kappa}{2}\int\limits_{0}^{\Le} dx_d \; \langle (\bpi \cdot \partial_{x_d}\bpi)^2 \rangle_> = -\frac{\kappa}{2}\int\limits_{0}^{\Le} dx_d \; \bigg[& (\bpi^< \cdot \partial_{x_d}\bpi^<)^2 + |\bpi^<|^2\frac{1}{(2\Le)^2}\sum_q^> q^2 G_{0q} \\& + |\partial_{x_d}\bpi^<|^2\frac{1}{(2\Le)^2}\sum_q^> G_{0q} + \langle (\bpi^> \cdot \partial_{x_d}\bpi^>)^2 \rangle_> \bigg],
\end{split}
\end{equation}
and
\begin{equation}
\begin{split}
\frac{g}{8}\iint\limits_{0}^{\Le} dx_d dx'_d \; \langle |\bpi(x_d)|^2|\bpi(x'_d)|^2 \rangle_> &= \frac{g}{8}\iint\limits_{0}^{\Le} dx_d dx'_d \; \bigg[ |\bpi^<(x_d)|^2|\bpi^<(x'_d)|^2 + \langle |\bpi_>(x_d)|^2|\bpi_>(x'_d)|^2 \rangle_> \bigg] \\
& \hspace{12pt} + \frac{g}{8}\int\limits_{0}^{\Le} dx_d \; 2(d-1)|\bpi^<(x_d)|^2 \frac{1}{(2\Le)}\sum_q^> G_{0q}.
\end{split}
\end{equation}
In these equations,
\begin{equation}
G_{0q} = \frac{4L_0T}{\tau + \kappa q^2}.
\end{equation}
Using the notation of Eq.~\eqref{EQ:dimlessquants}, we can also write
\begin{equation}
\frac{1}{4L_0^2}\sum_q^> G_{0q} = 2\bar{T}\mathcal{A}(\bar{\tau}).
\end{equation}
Feynman diagrams corresponding to these terms are included in Fig.~\ref{FIG:Diagrams}.

\begin{figure}
	\centering
		\includegraphics[width=.85\textwidth]{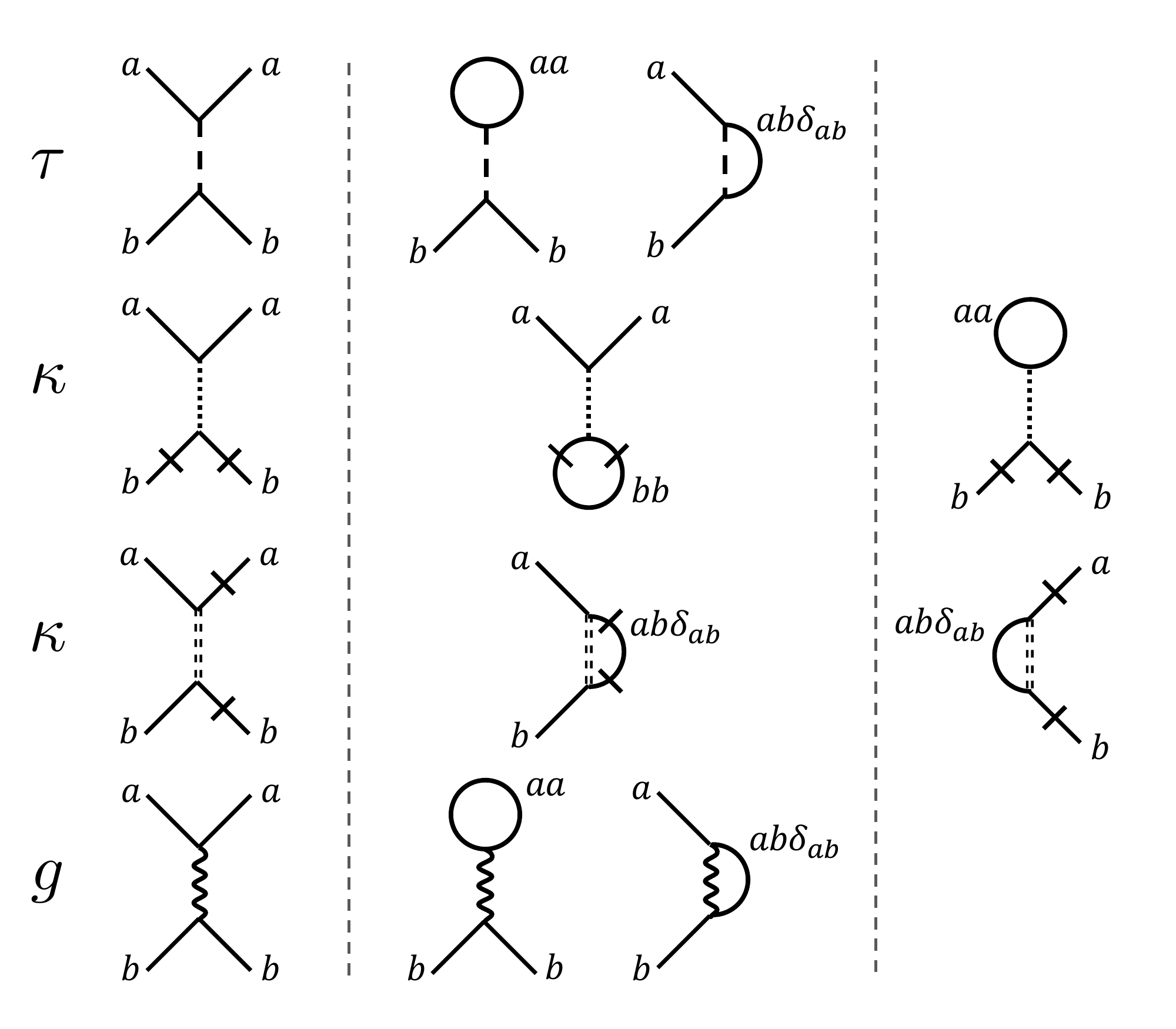}
	\caption{Feynman diagrams corresponding to terms in $\langle H_4 \rangle_>$. The diagrams are systematically divided into rows and columns: each row is associated with a single elastic parameter that is the coefficient of the originating term in $\langle H_4 \rangle_>$, while the columns specify which elastic parameter's renormalization the diagrams contribute to. Namely, the first column presents the basic vertex diagrams, while the second and third columns list those that renormalize $\tau$ and $\kappa$, respectively. Each external leg corresponds to a $\bpi^<$ field, and slashes denote spatial derivatives with respect to $x_d$. The various internal lines differentiate between the interactions and are used for index bookkeeping.}
	\label{FIG:Diagrams}
\end{figure}

As mentioned previously, we are interested in extracting the contribution to the coefficients of the quadratic-order $\bpi^<$ terms from $\left< H_4\right>_>$. Collecting terms, and defining renormalized elastic parameters $\tilde\t$ and $\tilde\br$ as the modified coefficients, we find
\begin{align}
\tilde{\tau} &= \tau + \frac{1}{2}\left[-(d+1)\tau + (d-1)gL_0 \right]\frac{1}{(2\Le)^2}\sum_{q}\limits^> G_{0q} - \kappa \left[1 + \frac{3}{2}(d-1)\right]\frac{1}{(2\Le)^2}\sum_{q}\limits^> q^2 G_{0q}, \\
\tilde{\kappa} &= \kappa\left\{1 - \left[1 + \frac{3}{2}(d-1)\right]\frac{1}{(2\Le)^2}\sum_{q}\limits^> G_{0q}\right\}.
\end{align}
The $\frac{1}{(2\Le)^2}\sum_{q}\limits^> q^2 G_{0q}$ term appears to have an ultraviolet divergence, but it actually vanishes. This is because it originates from quartic-order terms in the bending energy (see Eq.~\eqref{EQ:H4}) where the spatial derivatives are on the legs that combine to form the loops in the Feynman diagrams. This corresponds to a factor of $|\bpi'|^2$, which is the leading-order term in the gradient expansion of the difference in orientation between neighboring segments on the rod.  We can show this by restoring the full form of this factor for a segmented rod, $\sum_{x_d}|\bpi(x_d) - \bpi(x_d+\ell_0)|^2$, where $\ell_0$ is the projected length of each segment, and writing it in momentum space. Doing so, we obtain a factor of $1-\cos(q\ell_0)$ rather than only the leading-order term $q^2$. Here, $q = n\pi/N\ell_0$, so that $q\ell_0 = n\pi/N$. 
Taking the continuum limit where $N \rightarrow \infty$, the factor $1-\cos(q\ell_0)$ is highly oscillatory and 
the thus the whole expression, $\frac{1}{(2\Le)^2}\sum_{q}\limits^> [1-\cos(q\ell_0)] G_{0q}$ vanishes.

Simplifying these equations, we obtain the expressions for the renormalized elastic parameters, $\tilde{\tau}$ and $\tilde{\kappa}$, in Eq.~\eqref{EQ:Tilde}.
\end{widetext}

\section{Deriving the effective force}\label{APP:EForce}
In this section, we derive the effective force as perscribed in Eq.~\eqref{EQ:dFdL}. Starting from Eq.~\eqref{EQ:Fl} and building on the calculations of App.~\ref{APP:Reno}, we have that
\begin{align}
F^<(\bpi^<) &= H_0 + H_2(\bpi^<) + \langle H_4 \rangle_> - T\ln Z_0^> \\
&= H_0 + F_2^<(\bpi^<) + F_4^<(\bpi^<) - T\ln Z_0^> + O(T^2), \nonumber
\end{align}
with $F_2^<(\bpi^<)$ defined as in Eq.~\eqref{EQ:F2}, and $F_4^<(\bpi^<)$ containing terms quartic order in $\bpi^<$. The $O(T^2)$ terms, arising from 4-point correlation functions of $\bpi^>$, can be discarded. It is more convenient, going forward, to write $F_2^<$ and $F_4^<$ in terms of $\bpi_1$:
\begin{equation}
F_2^< = \frac{\tilde{\t} + \tilde{\br}\frac{\pi^2}{\Le^2}}{4\Le}|\bpi_1|^2 \equiv b_2T|\bpi_1|^2
\end{equation}
and
\begin{align}
F_4^< &= \frac{-\frac{3}{2}\t +5\t_c(0) + \sr\Le}{32 \Le^3} |\bpi_1|^4 \nonumber \\
&\approx \frac{\sr}{32 \Le^2}|\bpi_1|^4 \equiv b_4T|\bpi_1|^4,
\end{align}
where 
\begin{align}
	b_2 &\equiv \frac{\tilde{\t} + \tilde{\br}\frac{\pi^2}{\Le^2}}{4T\Le} \nonumber\\
	b_4 &\equiv 
	\frac{g}{32 T\Le^2}.   
\end{align}
Once again, we have taken the limit of the stretching stiffness much stronger than the bending stiffness in simplifying the expression for $F_4^<$. Thus, the Landau free energy becomes
\begin{equation}
F^< = H_0 - T\ln Z_0^> + b_2T|\bpi_1|^2 + b_4T|\bpi_1|^4.
\end{equation}
Since the first two terms are independent of $\bpi_1$, we can easily obtain an expression for the free energy,
\begin{equation}
F = H_0 - T\ln Z_0^> - T\ln \int_{-\infty}^{\infty} d^{d-1} \bpi_1 e^{-b_2 |\bpi_1|^2 -b_4 |\bpi_1|^4}.
\end{equation}
We now proceed to compute the latter two terms in this expression.

First,
\begin{align}
	-T\ln Z_0^> &= -T \ln \int \mathcal{D}\bpi^> e^{-H_2(\bpi^>)/T} \nonumber\\
	&= -T\ln \prod_{n=2}^{\infty} \left( \frac{4\pi T\Le}{ \t + \br (\pi n/\Le)^2} \right)^{\frac{d-1}{2}} \nonumber\\
	&= -\frac{T(d-1)}{2} \sum _{n=2}^{\infty} \ln \frac{4\pi T\Le}{ \t + \br (\pi n/\Le)^2} .
\end{align}
Second, we need to evaluate the integral 
\begin{align}
	Z_1 \equiv \int_{-\infty}^{\infty} d^{d-1} \bpi_1 e^{-b_2 |\bpi_1|^2 -b_4 |\bpi_1|^4}.
\end{align}
This integral can be evaluated using the $(d-1)$-dimensional spherical coordinates; we find
\begin{align}\label{EQ:IntI}
	Z_1
	& = \frac{1}{4}b_4^{-\frac{d-1}{4}} \Omega_{d-1}\Bigg[
		\Gamma\left(\frac{d-1}{4}\right) {}_1 F_1\left(\frac{d-1}{4};\frac{1}{2};\frac{c}{4}\right) \nonumber\\
		&\quad \mp\sqrt{c}\, \Gamma\left(\frac{d+1}{4}\right) {}_1 F_1\left(\frac{d+1}{4};\frac{3}{2};\frac{c}{4}\right)
	\Bigg],
\end{align}
where $\Omega_{d-1}$ is the solid angle subtended by the $(d-1)$-dimensional hypersphere, the dimensionless number
\begin{equation}
c\equiv \frac{|b_2|^2}{b_4} = \frac{2\left(\tilde{\tau} + \tilde{\kappa} \frac{\pi^2}{L_0^2} \right)^2}{gT},
\end{equation}
and ${}_1 F_1$ represents the Kummer confluent hypergeometric function.  The $-$ sign in Eq.~\eqref{EQ:IntI} applies to $b_2>0$, which is the straight phase, whereas the $+$ sign corresponds to $b_2<0$, the buckled phase.

To better understand the expression in Eq.~\eqref{EQ:IntI}, we expand it in different regimes.  The behavior of the ${}_1 F_1$ function takes different limits for $c\ll 1$ (close to the transition -- the \emph{critical} regime) and $c\gg 1$ (far from the transition).  The boundary between these two regimes, determined by $c \sim 1$, is indicated by the dashed curves in Figs.~\ref{FIG:PhaseDiagFull} and \ref{FIG:PhaseDiagrams}. For $c\ll 1$, we have
\begin{align}\label{EQ:ZCrit}
Z_1 \approx \frac{1}{4}b_4^{-\frac{d-1}{4}} \Omega_{d-1}\Gamma\left(\frac{d-1}{4}\right) \Bigg[1 - \tilde{\Gamma}\frac{b_2}{\sqrt{b_4}} + \frac{d-1}{8}\frac{|b_2|^2}{b_4} \Bigg],
\end{align}
where $\tilde{\Gamma} \equiv \Gamma(\tfrac{d+1}{4})/\Gamma(\tfrac{d-1}{4})$.
For $c\gg 1$, the asymptotic expressions depend on the phase of the rod: for the straight phase ($b_2 > 0$),
\begin{equation}\label{EQ:ZStraight}
Z_1 = \frac{1}{2}\Omega_{d-1}\Gamma\left(\frac{d-1}{2}\right)b_2^{-\frac{d-1}{2}},
\end{equation}
while for the buckled phase ($b_2 < 0$),
\begin{equation}\label{EQ:ZBuckled}
Z_1 = \frac{\sqrt{\pi}}{2^{\frac{d-1}{2}}}\Omega_{d-1}\left|b_2\right|^{\frac{d-3}{2}} b_4^{-\frac{d-2}{2}} e^{c/4}.
\end{equation}
The expressions for $Z_1$ in the $c \gg 1$ regime yield simple expressions when specializing to $d=2$ and $d=3$, so it is useful to explicitly list them:
\begin{equation}\label{EQ:ZNonCrit}
	Z_1 =
	\begin{cases}
		\sqrt{\pi/b_2} & b_2>0 \textrm{ and } d=2, \\
		\pi/b_2  & b_2>0 \textrm{ and } d=3, \\
		\sqrt{2\pi/|b_2|} \, e^{c/4} & b_2<0 \textrm{ and } d=2,\\
		\sqrt{\pi^3/b_4} \, e^{c/4} & b_2<0 \textrm{ and } d=3.\\
	\end{cases}
\end{equation}
The $e^{c/4}$ factor in the latter two equations comes from the finite expectation value of $\bpi_1$ when $b_2<0$.  It is straightforward to see this by plugging $\hat{\bpi}_1$ -- as given in Eq.~\eqref{EQ:pistar} -- into $H$.  

Next, we put the terms together and derive the effective force.  Following Eq.~\eqref{EQ:dFdL},
\begin{align}
	f &= \frac{\partial H_0}{\partial \Le} - Tg \frac{\partial }{\partial \t} \ln Z_0^> -Tg \frac{\partial }{\partial \t} \ln Z_1
	\nonumber\\
	&= \t + \frac{d-1}{2} \bar{g}\bar{T}|\t_c(0)|\mathcal{A}(\bar{\t}) + f_1,
\end{align}
where $f_1$ is from the $\ln Z_1$ term and can be expanded in the various limits.  

In the critical regime, we use Eq.~\eqref{EQ:ZCrit} and find
\begin{align}\label{EQ:f1Crit}
f_1 ={}& \left(\tilde{\Gamma}\sqrt{2gT} + \frac{1-d +  4\tilde{\Gamma}^2}{2} \left(\tilde{\tau} + \tilde{\br}\frac{\pi^2}{\Le^2}\right) \right) \nonumber \\
& \;\; \times \left(1 - (d-1)\bar{g}\bar{T}\mathcal{A}'(\bar\t) \right),
\end{align}
where $\mathcal{A}'(\bar\t)$ is the derivative of $\mathcal{A}(\bar\t)$ with respect to $\bar\t$. Deep in the straight phase, we use Eq.~\eqref{EQ:ZStraight} to obtain
\begin{align}\label{EQ:fS}
	f_1 = \frac{d-1}{2} \frac{gT}{\tilde{\t}+\tilde{\br}(\pi/\Le)^2} \left(1 - (d-1)\bar{g}\bar{T}\mathcal{A}'(\bar\t) \right),
\end{align}
while deep in the buckled phase, Eq.~\eqref{EQ:ZBuckled} gives us
\begin{align}\label{EQ:fB}
f_1 ={}& \left( -\tilde{\t} - \tilde{\br}\frac{\pi^2}{\Le^2} + \frac{3-d}{2} \frac{gT}{\tilde{\t} + \tilde{\br}(\pi/\Le)^2} \right) \nonumber \\
& \;\; \times \left(1 - (d-1)\bar{g}\bar{T}\mathcal{A}'(\bar\t) \right).
\end{align}

In the latter regimes, where $c \gg 1$, it turns out that $\bar{g}\bar{T} \ll 1$; therefore, we can write the expressions for $f_1$ to $O(\bar{g}\bar{T})$. The complete force expressions then simply become a leading-order term plus an $O(\bar{g}\bar{T})$ correction. Specifically, in the straight phase,
\begin{align}
f_1 = \frac{d-1}{2}\bar{g}\bar{T}|\t_c(0)|\frac{1}{1-\bar\t},
\end{align}
so that
\begin{align}
f = \t + \frac{d-1}{2}\bar{g}\bar{T}|\t_c(0)|\left[\mathcal{A}(\bar{\t}) + \frac{1}{1-\bar\t} \right],
\end{align}
and in the buckled phase,
\begin{align}
f_1 ={}& \t_c(0) - \t - \bar{g}\bar{T}|\t_c(0)| \Bigg[(d-1)\mathcal{A}'(\bar{\t})(\bar\t - 1) \nonumber \\
& \hspace{48pt} + (d-1)\mathcal{A}(\bar{\t}) + \frac{3-d}{2}\frac{1}{\bar\t - 1} \Bigg],
\end{align}
so that
\begin{align}
f ={}& \t_c(0) - \bar{g}\bar{T}|\t_c(0)| \Bigg[(d-1)\mathcal{A}'(\bar{\t})(\bar\t - 1) \nonumber \\
& \hspace{48pt} + \frac{d-1}{2}\mathcal{A}(\bar{\t}) + \frac{3-d}{2}\frac{1}{\bar\t - 1} \Bigg].
\end{align}
Notice the major difference between the two final expressions for the effective force: deep in the straight phase, the force is just the original/unmodified compression with a small $O(\bar{g}\bar{T})$ correction; on the other hand, deep in the buckled phase, the force is the zero-temperature critical compression with a small correction of the same order.

The critical regime, however, is not constrained to only small values of $\bar{g}\bar{T}$, so a similar expansion cannot be made everywhere; therefore, we further divide this regime into two limiting cases. In the region where $\bar{g}\bar{T} \ll 1$, Eq.~\eqref{EQ:f1Crit} becomes
\begin{equation}
f_1 = \tilde{\Gamma}\sqrt{2gT} + (\t - \t_c(T))\frac{1-d+4\tilde{\Gamma}^2}{2},
\end{equation}
where we discard all corrections of $O(\bar{g}\bar{T})$ and also note that $\t - \t_c(0) = \t - \t_c(T) + O(\bar\sr\bar{T})$, using Eq.~\eqref{EQ:tauCT}. Furthermore, in this regime, $\tau$ will deviate minimally from $\tau_c(T)$; therefore, we can simply take $f_1 \approx \tilde{\Gamma}\sqrt{2gT}$ -- which is indeed the value of $f_1$ on the transition curve -- as a reasonable approximation for the entire critical region (for $\bar{g}\bar{T} \ll 1$). Thus, the total force in this regime is
\begin{align}
f = \tau + \tilde{\Gamma}\sqrt{2gT},
\end{align}
which indicates an $O(\sqrt{T})$ correction to the force in the critical regime.

Finally, when $\bar{g}\bar{T} \sim O(1)$ in the critical regime, the $f_1$ contribution to the total force is suppressed, as $1 - (d-1)\bar{g}\bar{T}\mathcal{A}'(\bar{\t}) \approx 0$. In this case,
\begin{align}
f = \tau + \frac{d-1}{2}\bar{g}\bar{T}|\t_c(0)|\mathcal{A}(\bar{\t}),
\end{align}
and there is, once again, an $O(\bar{g}\bar{T})$ correction to the compression.


\end{document}